\documentclass[aps,prl,twocolumn,superscriptaddress,showpacs,preprintnumbers,amsmath,amssymb]{revtex4-2}

\usepackage{comment}
\usepackage{orcidlink}

\usepackage{epsfig}
\usepackage{bm}
\usepackage{color}
\usepackage{overpic}
\usepackage{siunitx}
\usepackage{tabularx}

\usepackage{makecell}
\usepackage{graphicx}
\usepackage{lineno}

\usepackage{subfigure}
\usepackage{overpic}
\usepackage{epsfig}
\usepackage{dcolumn}
\usepackage{bm}
\usepackage{color}
\usepackage{siunitx}
\usepackage{ulem}

\newcommand{\yones}{\Upsilon(1S)}
\newcommand{\ytwos}{\Upsilon(2S)}
\newcommand{\yonetwos}{\Upsilon(1S,2S)}

\newcommand{\eff}{\varepsilon}
\newcommand{\BR}{{\cal B}}

\newcommand{\pim}{\pi^-}

\newcommand{\jpsillb}{\jpsi\,\Lambda/\bar{\Lambda}}
\newcommand{\kl}{K_L}

\newcommand{\jpsi}{J/\psi}

\newcommand{\pc}{P_{c\bar{c}}^{+}}
\newcommand{\pcs}{P_{c\bar{c}s}^{0}}
\newcommand{\pcsbar}{\bar{P}_{c\bar{c}s}^{0}}
\newcommand{\pcsa}{P_{c\bar{c}s}(4459)^{0}}
\newcommand{\pcsb}{P_{c\bar{c}s}(4338)^{0}}
\newcommand{\pcsabar}{\bar{P}_{c\bar{c}s}(4459)^{0}}
\newcommand{\pcsbbar}{\bar{P}_{c\bar{c}s}(4338)^{0}}

\newcommand{\pjpsi}{pJ/\psi}
\newcommand{\lamjpsi}{J/\psi\Lambda}

\newcommand{\EE}{e^+e^-}
\newcommand{\MM}{\mu^+\mu^-}
\newcommand{\LL}{\ell^+\ell^-}

\newcommand{\ccb}{c\bar{c}}
\newcommand{\qqb}{q\bar{q}}
\newcommand{\ppi}{p\pi^-}

\newcommand{\bcntr}{\begin{center}}
\newcommand{\ecntr}{\end{center}}
\newcommand{\beq}{\begin{equation}}
\newcommand{\eeq}{\end{equation}}
\newcommand{\beqar}{\begin{eqnarray}}
\newcommand{\eeqar}{\end{eqnarray}}
\newcommand{\bitm}{\begin{itemize}}

\newcommand{\bitmb}{\begin{itemize}}
\newcommand{\benub}{\begin{enumerate}}
\newcommand{\eitm}{\end{itemize}}

\newcommand{\bct}{\begin{center}}
\newcommand{\ect}{\end{center}}
\newcommand{\bpic}{\begin{overpic}}
\newcommand{\epic}{\end{overpic}}

\newcommand{\cm}{\si{\centi\metre}}

\newcommand{\fb}{\si{\femto\barn}}

\newcommand{\infb}{\fb^{-1}}

\newcommand{\gevcs}{\hbox{GeV}/c^2}

\newcommand{\gev}{\hbox{GeV}}
\newcommand{\mevcs}{\hbox{MeV}/c^2}

\newcommand{\mev}{\hbox{MeV}}

\newcommand{\reduline}{\bgroup\markoverwith
{\textcolor{red}{\rule[0.5ex]{2pt}{0.4pt}}}\ULon}

\def\Journal#1#2#3#4{{#1} {\bf #2}, #3 (#4)}

\def\NIMA{Nucl. Instrum. Methods A}
\def\NIMPRA{Nucl. Instrum. Methods Phys. Res., Sect. A}

\def\PL{Phys. Lett.}
\def\PLB{Phys. Lett. B}
\def\PRL{Phys. Rev. Lett.}
\def\PRD{Phys. Rev. D}
\def\PRP{Phys. Rep.}
\def\CPC{Chin. Phys. C}

\def\EPJC{Eur. Phys. J. C}

\def\PTEP{Prog. Theor. Exp. Phys. }
\def\ARNPS{Annu. Rev. Nucl. Part. Sci.}

\def\SciBU{Sci. Bull.}

\def\FBS{Few Body Syst.}
\def\pop{Progr. Phys.}
\def\rmp{Rev. Mod. Phys.}
\def\ppnp{Prog. Part. Nucl. Phys.}

\begin{document}


\title{
   \quad\\[1.0cm]
   Search
for $\pcsa$ and $\pcsb$ in $\yonetwos$ inclusive decays at Belle}



  \author{I.~Adachi\,\orcidlink{0000-0003-2287-0173}} 
  \author{L.~Aggarwal\,\orcidlink{0000-0002-0909-7537}} 
  \author{H.~Ahmed\,\orcidlink{0000-0003-3976-7498}} 
  \author{J.~K.~Ahn\,\orcidlink{0000-0002-5795-2243}} 
  \author{H.~Aihara\,\orcidlink{0000-0002-1907-5964}} 
  \author{N.~Akopov\,\orcidlink{0000-0002-4425-2096}} 
  \author{M.~Alhakami\,\orcidlink{0000-0002-2234-8628}} 
  \author{A.~Aloisio\,\orcidlink{0000-0002-3883-6693}} 
  \author{N.~Althubiti\,\orcidlink{0000-0003-1513-0409}} 
  \author{D.~M.~Asner\,\orcidlink{0000-0002-1586-5790}} 
% ...
\collaboration{The Belle and Belle II Collaborations}

\begin{abstract}
Using data samples of 102 million $\yones$ events and 158 million $\ytwos$ events collected by
the Belle detector at the KEKB asymmetric-energy $\EE$ collider, we search for $[udsc\bar{c}]$
pentaquark states decaying to $\lamjpsi$. Using the first observations of $\yonetwos$ inclusive
decays to $\lamjpsi$, we find evidence of the $\pcsa$ state with a local significance of 3.3 standard
deviations, including statistical and systematic uncertainties. We measure the mass and width
of the $\pcsa$ to be $(4471.7 \pm 4.8 \pm 0.6)~\mevcs$ and $(22 \pm 13 \pm 3)~\mev$,
respectively. The branching fractions for $\pcsa$ production are measured to be $\BR[\yones\to
\pcsa/\pcsabar + anything] = (3.5 \pm 2.0 \pm 0.2) \times 10^{-6}$ and $\BR[\ytwos \to
\pcsa/\pcsabar + anything] = (2.9 \pm 1.7 \pm 0.4) \times 10^{-6}$. The inclusive branching
fractions of $\yonetwos \to \jpsillb$ are measured to be  $\BR[\yones \to \jpsillb + anything]
= ( 36.9 \pm 5.3 \pm 2.4) \times 10^{-6}$ and $\BR[\ytwos \to \jpsillb + anything ] = (22.3
\pm 5.7 \pm 3.1) \times 10^{-6}$. We measure the visible cross section $\sigma(e^+ e^- \to
\jpsillb + anything ) = (90 \pm 14 \pm 6)~\fb$ for the continuum production at $\sqrt{s} =
10.52~\gev$. In all cases, the first uncertainties are statistical and the second are
systematic.
\end{abstract}


\maketitle

Interest in pentaquark states started in the 1960s as both Gell-Mann and Zweig postulated
their existence in their first descriptions of the quark model~\cite{quark-model, quark-model2}.
The first observation of the charged pentaquark state candidates, $\pc$, with valence quark
content [$uud\ccb$] was reported in the decay $\Lambda_b \to \jpsi pK^-$ by the LHCb
experiment~\cite{pc,lhcb_pc_2019}. In a subsequent search for a neutral pentaquark, LHCb
reported evidence ($3.1\sigma$) of a pentaquark candidate state with a suggested quark
assignment $[uds\ccb]$~\cite{th}, named the $\pcsa$ with a mass of $(4458.8 \pm
2.9_{-1.1}^{+4.7})~\mevcs$ and a width of $(17.3 \pm 6.5_{-5.7}^{+8.0})~\mev$,
in the $\lamjpsi$ substructure of the decay $\Xi_b^- \to \jpsi\Lambda K^-$~\cite{pcs}. Here and
hereinafter, the first uncertainty quoted is statistical, and the second is systematic. Another
candidate $\pcsb$, sharing the same suggested valence quark content, was discovered in the decay
of $B^- \to \jpsi \Lambda \bar{p}$~\cite{pcs4338} with a statistical significance exceeding
$15\sigma$, a measured mass of $(4338.2 \pm 0.7 \pm 0.4)~\mevcs$, and a width of $(7.0 \pm 1.2
\pm 1.3)~\mev$.

The pentaquark candidates found by LHCb are all at masses close to the production thresholds
of ordinary baryon-meson states, i.e., $\Sigma_{c}^{+}\bar{D}^{(*)0}$ for the $\pc$
states~\cite{pc,lhcb_pc_2019}, $\Xi_{c}^{0}\bar{D}^{*0}$ for the $\pcsa$ state~\cite{pcs} and
$\Xi_{c}^{+}D^{-}$ for the $\pcsb$ state~\cite{pcs4338}. There are various interpretations of
these states, including tightly-bound pentaquark states~\cite{bound_1, bound_2}, loosely-bound
baryon-meson molecular states~\cite{mol_1,mol_2}, or the product of rescattering
effects~\cite{resca}. However, their nature is still largely unknown, and further investigation
is needed. Moreover, these states have so far only been reported by LHCb, and it is essential
to provide independent confirmation of their existence.

Theoretical considerations suggest that $\yones$ and $\ytwos$ decays could produce final
states of matter with unusual quark configurations~\cite{Rosner}. Meanwhile, the observations
of inclusive production of the antideuteron, a candidate for a hexaquark bound
system~\cite{deuteron}, by the ARGUS, CLEO, and BaBar experiments in $\yones$ and $\ytwos$
inclusive decays~\cite{cleo_db_yns, ARGUS_db_yns, Babar_db_yns}, suggest searching for a $\pc$
or $\pcs$ state in the same data sample. In a study of the $\pjpsi$ final state from $\yonetwos$
inclusive decays~\cite{dongxu}, Belle saw no significant $\pc$ signal. However, this work did
report a branching fraction $\BR[\yonetwos \to \jpsi\,p/\bar{p} + anything]$ at the $10^{-5}$
level.

This Letter reports the results of a search for $\pcs$ states in the $\lamjpsi$ final state of
$\yonetwos$ inclusive decays using the world's largest $\yones$ and $\ytwos$ data samples,
produced by the KEKB collider~\cite{KEKB1, KEKB2} and collected by the Belle
detector~\cite{Belle}. Here, the $\jpsi$ is reconstructed in the $\LL$ ($l = e,~\mu$) final
state and $\Lambda$ in its decay to $p\pim$. Inclusion of charge-conjugate processes is implied.
The $\yones$ data sample has an integrated luminosity of $\mathcal{L}_{\yones} = 5.8~\infb$ and
$(1.02 \pm 0.02) \times 10^8$ $\yones$ events~\cite{y1sevnt}, and the $\ytwos$ data sample has
$\mathcal{L}_{\ytwos} = 24.5~\infb$ and $(1.58 \pm 0.04)\times 10^8$ $\ytwos$
events~\cite{y2sevnt}. As $\yonetwos$ are produced from $\EE$ annihilation, a data sample
collected at $\sqrt{s} = 10.52~\gev$ with an integrated luminosity of $\mathcal{L}_{\rm cont}
= 89~\infb$ (referred to as the ``continuum data sample") is used to study continuumbackground.


To optimize the signal selection criteria and determine the reconstruction efficiencies and
resolutions, we use EvtGen~\cite{evtgen} to simulate the signal Monte Carlo (MC) samples of
$\yonetwos \to \pcs \bar{\Lambda} + \qqb$ with $\pcs\to \lamjpsi$ based on phase
space~\cite{evtgen}. Here, $\pcs$ represents $\pcsa$ or $\pcsb$, and $\qqb~(q = u, d, s, c)$
denotes a quark-antiquark pair whose hadronization is simulated using PYTHIA~\cite{pythia}.
To investigate the efficiency and the resolution dependence on the $\lamjpsi$ invariant mass
($M_{\lamjpsi}$), we generate a range of signal MC samples with $M_{\lamjpsi}$ varying from
$4.3~\gevcs$ to $5.6~\gevcs$ and an intrinsic width set to zero. To examine non-resonant
$\lamjpsi$, we generate MC samples of $\yonetwos\to \lamjpsi \bar{p} K^{+} + \qqb$ using phase
space~\cite{evtgen} and these are referred to as the ``no-$\pcs$ MC samples". The $\qqb$ system
in the $\yones$ [$\ytwos$] decays is generated with a mass of $2.6~\gevcs$ ($3.2~\gevcs$) and
a broad width of $1.7~\gev$. We employ a GEANT3-based MC technique~\cite{geant3} to simulate
the response of the Belle detector. We use the Belle II analysis software framework
(basf2)~\cite{basf2} to reconstruct the events for Belle data. The Belle data is converted to
the Belle II format for basf2 compatibility using the B2BII software package~\cite{b2bii}.


The charged tracks, except for those used in the reconstruction of $\Lambda$ candidates, are
selected to originate from the interaction point by requiring their impact parameters to be less
than $2.0~\cm$ along the beam direction ($dz$), and less than $0.2~\cm$ in the transverse plane
($dr$). Two tracks with opposite charges and a difference of $dz$ less than $0.2~\cm$ are
selected as candidates of the lepton pair from $\jpsi$ decay. Electrons are identified by having
$\mathcal{L}_e/(\mathcal{L}_e+\mathcal{L}_h) > 0.9$, where the electron likelihood
$\mathcal{L}_e$ and hadron likelihood $\mathcal{L}_h$ ($h=\pi, K, p$) are assigned based on
central drift chamber, aerogel threshold Cherenkov counter, and electromagnetic calorimeter
(ECL) information~\cite{pid, EID}. Tracks are identified as muons if they have
$\mathcal{L}_\mu/(\mathcal{L}_\mu + \mathcal{L}_\pi + \mathcal{L}_K) > 0.9$, where the muon
likelihood $\mathcal{L}_\mu$ is assigned based on the range and hit positions of the
extrapolated track in the $\kl$ and muon detector~\cite{MUID}. The particle identification
(PID) efficiency of a single lepton is $(93.9 \pm 0.2)\%$ for $e^\pm$ and $(91.9 \pm 0.2)\%$
for $\mu^\pm$. Bremsstrahlung photons detected in the ECL within a cone of 0.05 radians about
the original $e^\pm$ direction are incorporated into the calculation of the $\EE(\gamma)$
invariant mass. The $\Lambda$ candidates are reconstructed in the $\Lambda \to \ppi$ decay mode
using an artificial neural network~\cite{NN}, which uses vertex fit and the PID information,
$\mathcal{L}_{p}/(\mathcal{L}_{p} + \mathcal{L}_{\pi})$, to identify $p$ and $\pi^-$ candidates.

\begin{figure}[tbp]
\centering
\includegraphics[width=0.45\textwidth]{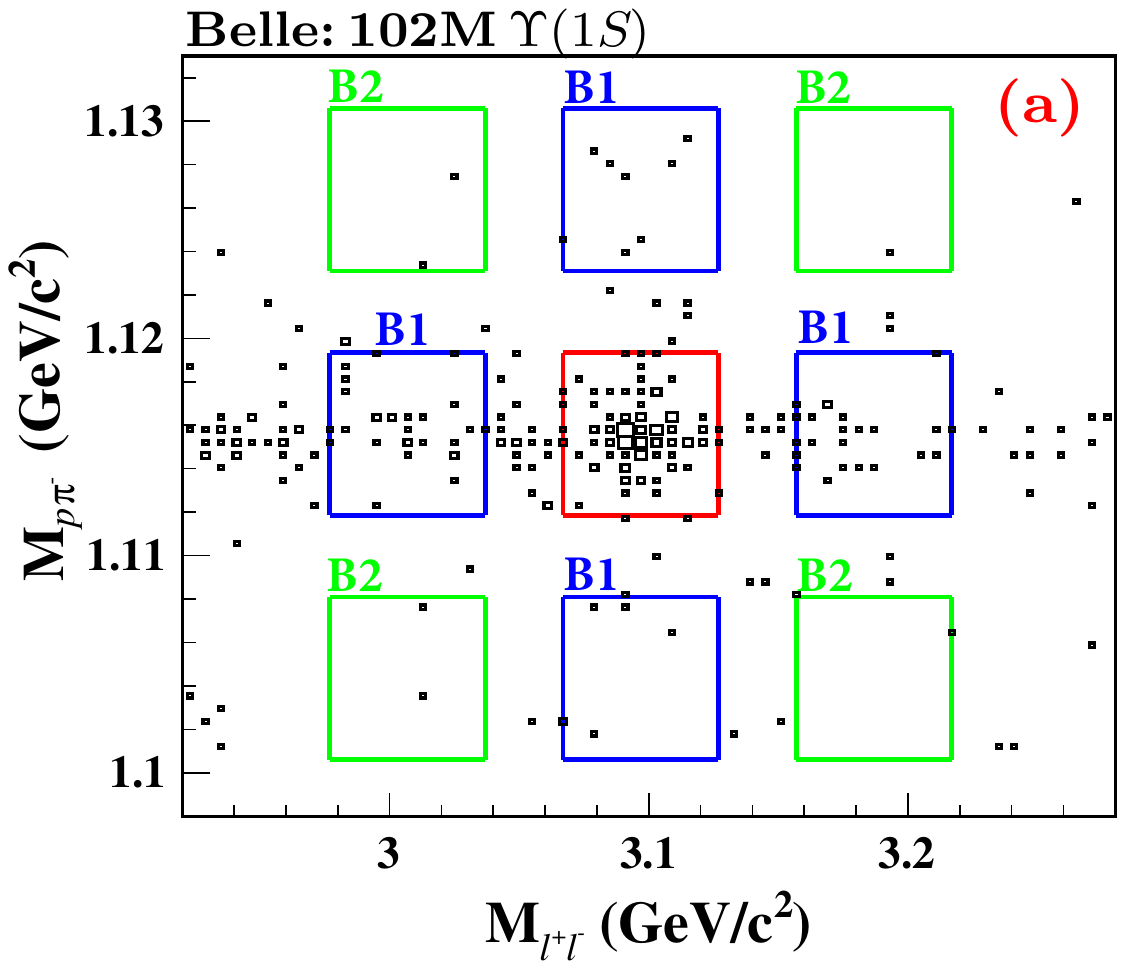}\\
\includegraphics[width=0.45\textwidth]{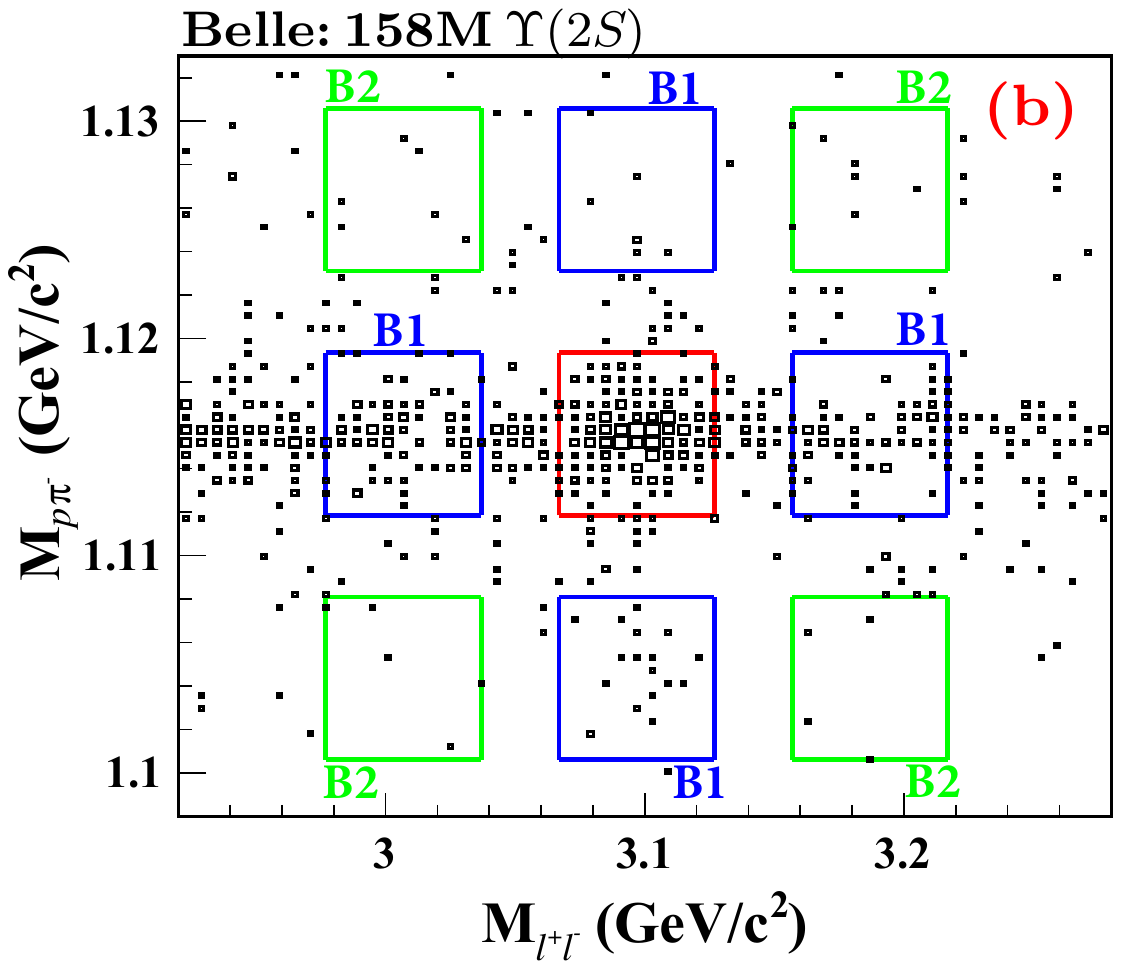}
\caption{Two-dimensional scatter plots of $M_{\LL}$ versus $M_{p\pi^-}$ from (a) the $\yones$
data sample and (b) the $\ytwos$ data sample. The central red boxes are the signal regions,
while the blue and green boxes around it are the two-dimensional sideband regions.}
\label{2d}
\end{figure}

Figure~\ref{2d} shows scatter plots of the invariant mass of the lepton pair ($M_{\LL}$) versus
the $p\pi^-$ pair ($M_{p\pi^-}$) from the $\yonetwos$ data samples. Clear $\jpsi$ and $\Lambda$
signals are visible. By fitting the $M_{\LL}$ distribution with a double Gaussian function for
the $\jpsi$ signals and a first-order Chebychev polynomial for the backgrounds, we obtain the
resolutions $\sigma_{\jpsi}^{\rm data} = 9.4 \pm 0.1~\mevcs$ in data and
$\sigma_{\jpsi}^{\rm MC} = 9.2 \pm 0.1~\mevcs$ in signal MC simulation. Similarly, we obtain
$\sigma_{\Lambda}^{\rm data} = 1.4 \pm 0.2~\mevcs$ and $\sigma_{\Lambda}^{\rm MC} = 1.3 \pm
0.1~\mevcs$ for the $\Lambda$ signals by fitting to the $M_{p\pi^-}$ distributions in data and
signal MC simulation. We define the signal regions as $|M_{\LL} - m_{\jpsi}| < 30~\mevcs$ and
$|M_{\ppi} - m_{\Lambda}| < 4.2~\mevcs$, where $m_{\jpsi}$ and $m_{\Lambda}$ are the nominal
masses of the $\jpsi$ and $\Lambda$~\cite{PDG}. The central red boxes in Fig.~\ref{2d} show
the signal regions. To estimate the backgrounds in the $\Lambda$ and $\jpsi$ reconstructions,
we define one-dimensional sideband regions as $|M_{\LL} - m_{\jpsi} \pm 90| < 30~\mevcs$ and
$|M_{\ppi} -m_{\Lambda} \pm 12.6| < 4.2~\mevcs$. Combining these one-dimensional mass ranges,
we define two-dimensional sideband regions shown by the rectangles in Fig.~\ref{2d} which
surround the signal area. There are three types of backgrounds in the signal region. The
backgrounds from true $\jpsi$ ($\Lambda$) and combinatorial $\LL$
($p\pi^-$) can be estimated using the yields in the two blue horizontal (vertical)
boxes scaled by a factor of 0.5. The backgrounds from combinatorial $p\pi^-$ and combinatorial $\LL$ can be estimated from the yields in the four green diagonal boxes scaled by a factor 0.25. It should be noted that when estimating the first two types, the third type is counted twice.  With $N_{B_1}$ ($N_{B_2}$) representing the sum of the events in
the four sideband regions nearest (diagonal to) the signal region, the yield of backgrounds to
the reconstructed $\Lambda$ and $\jpsi$ candidates is calculated to be
$N_B = 0.5  N_{B_1} - 0.25 N_{B_2}$.

After subtracting the backgrounds estimated from the two-dimensional sidebands, the numbers of
$\lamjpsi$ events in the signal regions are $N^{\lamjpsi}_{\yones} = 84 \pm 11$,
$N^{\lamjpsi}_{\ytwos} = 140 \pm 17$, and $N^{\lamjpsi}_{\rm cont} = 134 \pm 21$ in the
$\yones$, $\ytwos$, and continuum data samples, respectively. According to the no-$\pcs$
MC simulations, their efficiencies are $\eff_{\yones} = (26.5 \pm 0.2)\%$ and $\eff_{\ytwos}
= (27.0 \pm 0.2)\%$ in $\yones$ and $\ytwos$ inclusive decays, and $\eff_{\rm cont} =
(26.6 \pm 0.2)\%$ in the continuum process. To estimate the continuum background in the
$\yonetwos$ data samples from the continuum data sample, we scale the luminosity and correct
for the energy dependence of the cross section assuming $\sigma_{\EE} \propto 1/s$. This results in a scale factor $f_{\rm scale} = [\mathcal{L}_{\Upsilon}
\epsilon_{\Upsilon} s_{\rm cont}]/[\mathcal{L}_{\rm cont} \epsilon_{\rm cont} s_{\Upsilon}]
= 0.058$ and 0.266 for the $\yones$ and $\ytwos$ data samples, respectively.

We measure the cross section of the inclusive production of $\jpsillb$ in $\EE$ annihilation via
the equation
\begin{multline}
\sigma(\EE \to \jpsillb + anything ) = \\
\frac{N^{\lamjpsi}_{\rm cont}}{\mathcal{L}_{\rm cont}  \eff_{\rm cont} \BR(\jpsi\to \LL)
\BR(\Lambda \to p\pi^{-})  (1+\delta_{\rm ISR})},
\end{multline}
where $\BR(\jpsi\to \LL)$ and $\BR(\Lambda \to \ppi)$ are the world average values of the
branching fractions of $\jpsi$ and $\Lambda$ decays~\cite{PDG}, and the radiative correction
factor $1 + \delta_{\rm ISR}$ is calculated to be 0.82~\cite{isr01, isr02}. We obtain
$\sigma(\EE \to \jpsillb + anything ) = (90 \pm 14 \pm 6)~\fb$ at $\sqrt{s} = 10.52~\gev$,
where the systematic uncertainties are described below.

We then measure the production of $\jpsillb$ in $\yonetwos$ inclusive decays. We calculate the
branching fractions of $\yones$ inclusive decays to $\jpsillb$ using
\begin{multline}
\BR[\yones \to \jpsillb + anything] = \\
\frac{N^{\lamjpsi}_{\yones} - f_{\rm scale} N^{\lamjpsi}_{\rm cont}}
{N_{\yones} \eff_{\yones} \BR(\jpsi\to \LL) \BR(\Lambda \to p\pi^{-})},
\end{multline}
and find a value $(36.9 \pm 5.3 \pm 2.4) \times 10^{-6}$. Similarly, we obtain $\BR[\ytwos \to
\jpsillb + anything] = ( 32.0 \pm 5.5 \pm 3.0) \times 10^{-6}$. Subtracting the contribution due
to $\ytwos$ to $\yones$ transitions~\cite{PDG}, we find the branching fraction for direct
$\ytwos$ inclusive decays to be $\BR[\ytwos \to \jpsillb + anything] = (22.3 \pm 5.7 \pm 3.1)
\times 10^{-6}$. These are the first measurements of these inclusive branching fractions.

\begin{figure}[tbp]
\centering
\includegraphics[width=0.45\textwidth]{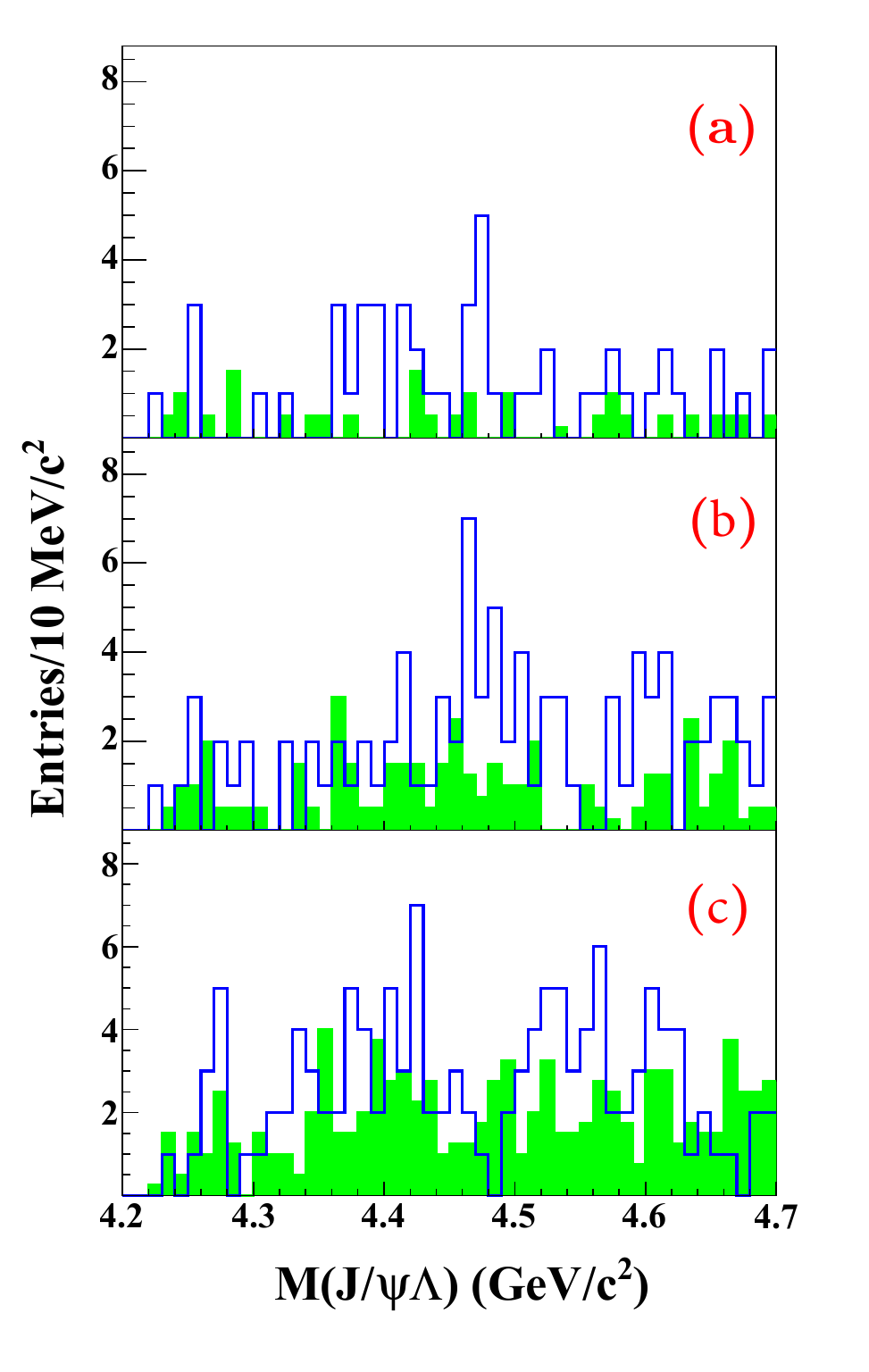}
\caption{The invariant mass distributions of $\lamjpsi$ from (a) the $\yones$ data sample,
(b) the $\ytwos$ data sample, and (c) the continuum data sample. The open blue blank histograms
show the events in signal regions, while the green filled histograms show the background
estimated from the two-dimensional sideband regions.}
\label{mlamjpsi}
\end{figure}

The $M_{\lamjpsi}$ distributions 
in the $\yones$, $\ytwos$ and continuum data samples are illustrated in Fig.~\ref{mlamjpsi}. To
avoid broadening due to the mass resolutions of the $\LL$ and $\ppi$ combinations, we use the
calculation $M_{\lamjpsi} = M_{\LL\ppi} - M_{\LL} - M_{\ppi} + m_{\jpsi} + m_{\Lambda}$, where
$M_{\LL\ppi}$ is the invariant mass calculated from the sum of the 4-momenta of the $\LL$ and
$\ppi$ pairs. From the signal MC, using this procedure improves the resolution in the
$M_{\lamjpsi}$ distribution to $2.2~\mevcs$ for $\pcsb$ and $2.8~\mevcs$ for $\pcsa$, compared
to about $11.5~\mevcs$ and $12.3~\mevcs$ in the $M_{\LL p\pi^{-}}$ distribution. As seen
in Fig.~\ref{mlamjpsi}, there are event accumulations near the mass of $\pcsa$ in the $\yones$
and $\ytwos$ data samples, but none in the $\pcsb$ region.

\begin{figure}[tbp]
\centering
\includegraphics[width=0.45\textwidth]{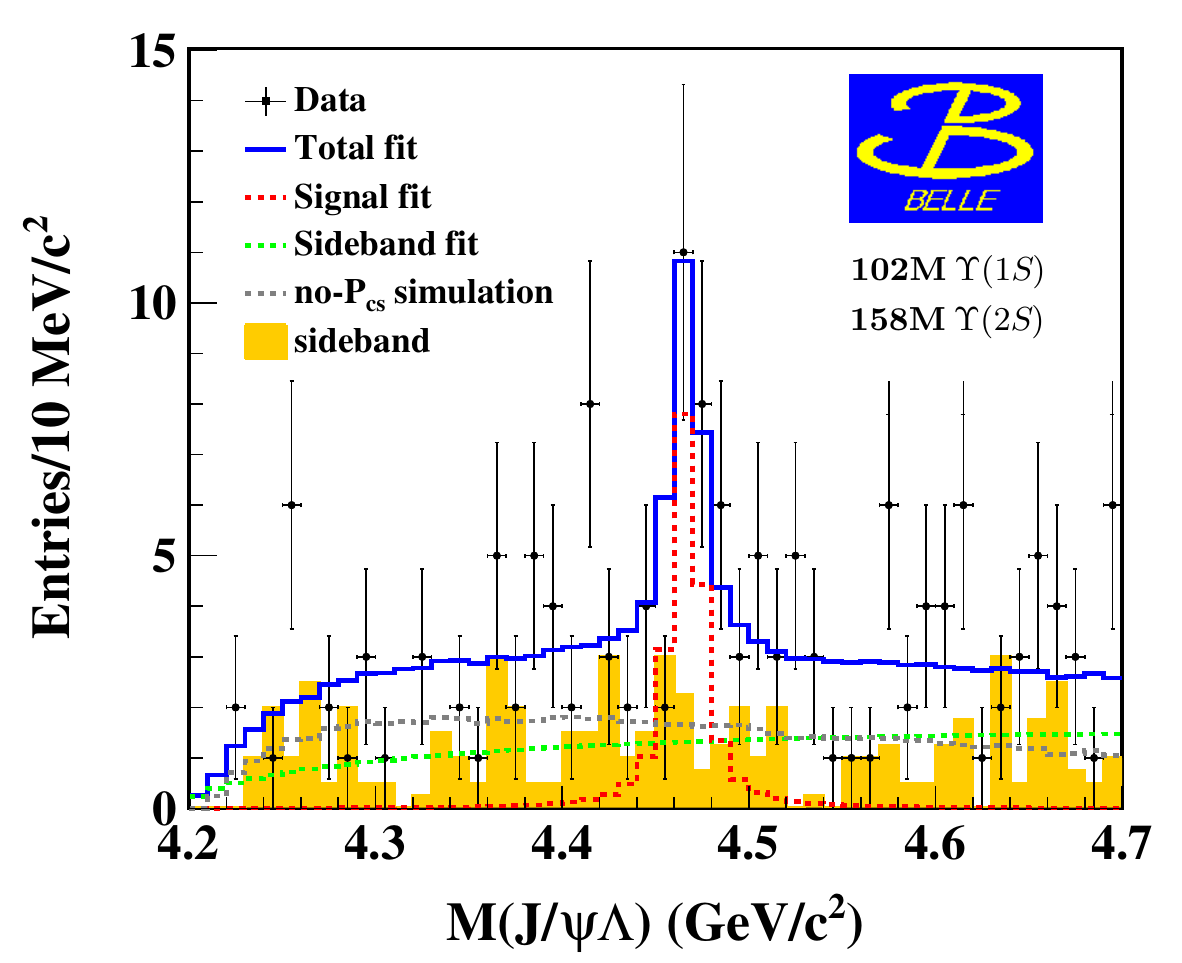}
\caption{The distribution of the invariant mass of $\lamjpsi$ in the combined $\yonetwos$ data
sample and the fit results with mass and width constrained. The points with error bars are the
data, and the yellow histogram is the background estimated from the two-dimensional sideband
regions. The solid curve shows the best fit results. The red dashed curve shows the signal. The
brown dashed curve shows the no-$\pcs$ component. The green dashed curve shows the fit to the
background estimated from the sidebands.}
\label{simfit}
\end{figure}

A binned maximum-likelihood fit is performed to the $\lamjpsi$ mass spectrum obtained from the
combined $\yonetwos$ data sample to study the excess, as shown in Fig.~\ref{simfit}.
The fit is performed with a bin width of $1~\mevcs$ rather than the $10~\mevcs$ used in the
figure for display purposes. We use this narrow bin width when finding our nominal fit as it retains the most information and is a close approximation to an unbinned fit. The found local significance is lower for this fit than those with larger bin widths. The probability density function~(PDF) used to describe the events
selected from the signal region is
\beq\label{eq-1}
f_{\rm PDF} = f_{\rm R}(m, \Gamma) + f_{\rm noPcs} +
f_{\rm SB}(m, c_0),
\eeq
where $f_{\rm R}$ is the PDF for the resonance, $f_{\rm SB}$ is the background
PDF estimated from the two-dimensional sidebands, and $f_{\rm noPcs}$ is for the
no-$\pcs$ production. Here, $f_{\rm R}$ is the convolution of a Breit-Wigner function and a
Gaussian function with the resolution fixed to the value of $2.8~\mevcs$ obtained from the
signal MC simulation. We use $\sqrt{M_{\lamjpsi} - M_{\rm thr}} \times e^{c_0 M_{\lamjpsi}}$
for $f_{\rm SB}$, where $M_{\rm thr} = 4.21~\gevcs$ is the mass threshold of $\lamjpsi$, and
$c_0$ is a coefficient determined by the fit. The PDF $f_{\rm noPcs}$ is the histogram
\sout{PDF} of $M_{\lamjpsi}$ obtained from the no-$\pcs$ MC simulation. We fit simultaneously
to the events from the signal region with $f_{\rm PDF}$ and events from the two-dimensional
sideband regions with $f_{\rm SB}$. The likelihood for the fit is denoted as $\mathcal{L}$.

Since the excess is close to the mass of $\pcsa$, we include a Gaussian constraint using prior
knowledge of the LHCb measurement~\cite{pcs} in the fit and minimize the value of
\beq \label{eq-2}
-2\ln\mathcal{L}^\prime \equiv -2\ln\mathcal{L} + \frac{(m-m_0)^2}{\sigma_{m_0}^2} +
\frac{(\Gamma-\Gamma_0)^2}{\sigma_{\Gamma_0}^2},
\eeq
where $m$ and $\Gamma$ are the mass and width of the signal structure, $m_0$ and $\Gamma_0$ are
the mean values from the LHCb measurement~\cite{pcs}, and $\sigma_{m_0}$ and $\sigma_{\Gamma_0}$
are their asymmetric uncertainties on $m_0$ and $\Gamma_0$. If the values of masses or
widths from the fits are greater than LHCb's measurements, positive uncertainties are quoted;
otherwise, negative uncertainties are used. The fit yields the number of $\pcsa$ signal events
$N_{\pcsa} = 21 \pm 5$. By removing the $f_{\rm R}$ term from $f_{\rm PDF}$ in Eq.~(\ref{eq-1}),
i.e., if we use the background-only hypothesis, the new fit yields a change
$\Delta(-2\ln\mathcal{L}^\prime) = 13.01$. The significance of the signal is estimated using a
pseudo-experiment technique. The pseudo-experiments are generated based on the fit result of
the background-only hypothesis, assuming a Poisson distribution of events in each bin of the
$M_{\lamjpsi}$ distribution. The fit in each pseudo-experiment follows the same procedures as
in the nominal fit. Among the $4.3 \times 10^5$ pseudo-experiments, only 160 have
$\Delta(-2\ln\mathcal{L}^\prime) > 13.01$. This corresponds to a $p$-value of
$3.8\times10^{-4}$ and thus a significance of $3.4\sigma$ for the $\pcsa$ in the combined
$\yonetwos$ data sample. To estimate the systematic uncertainty due to the background modeling,
we replace the exponential function in $f_{\rm SB}$ with a second-order Chebyshev polynomial.
This results in a significance for the $\pcsa$ of $3.3\sigma$ including systematic uncertainties.

We also perform a fit without the mass and width constraints. The fit yields a mass of $M_R =
(4471.7 \pm 4.8 \pm 0.6)~\mevcs$ and a width of $\Gamma_R = (22 \pm 13 \pm 3)~\mev$ for
this structure, where the systematic uncertainties are described below.
The fit results are consistent with the measurements for the $\pcsa$ as reported by LHCb, with differences of
$1.8\sigma$ for the mass and $0.3\sigma$ for the width. The local significance
is calculated to be $3.8\sigma$ given the change $\Delta(-2\ln\mathcal{L}) = 14.58$. In addition, we use a pseudo-experiment technique to find a global significance of the peak without any constraints of 2.8$\sigma$.

We also calculate the branching fraction for $\pcs$ production in the $\yonetwos$ inclusive
decays by
\begin{multline}
\label{brpcs}
\BR[\yonetwos \to \pcs/\pcsbar +anything] \BR(\pcs \to \lamjpsi) = \\
   \frac{N^{\rm fit}_{\pcs}} {N_{\yonetwos}
   \eff_{\pcs} \BR(\jpsi\to \LL) \BR(\Lambda \to p\pi^{-})},
\end{multline}
where $N^{\rm fit}_{\pcs}$ is the number of the signal events of $\pcs$ in the fit with
constrains, and $\eff_{\pcs}$ is the corresponding efficiency. Table~\ref{ul} shows the results
for $\pcsa$. In calculating the branching fraction in $\ytwos$ inclusive decays, the
contribution of the transition from $\ytwos$ to $\yones$ is removed.

We perform a fit to $M_{\lamjpsi}$ with both the $\pcsa$ and $\pcsb$ resonance included in
$f_{\rm R}$. This fit uses a histogram PDF obtained from a signal MC simulation of $\pcsb$.
Since there is no significant $\pcsb$ signal, we calculate the upper limits on the signal
yields [$N^{\rm fit, UL}_{\pcsb}$] at 90\% confidence level~(C.L.) by solving the equation
\beq
\label{eq-3}
\int_{0}^{N^{\rm fit, UL}_{\pcsb}} \mathcal{L}(N) dN/\int_{0}^{+\infty} \mathcal{L}(N) dN
= 0.90,
\eeq
where $N$ is the assumed signal yield, and $\mathcal{L}(N)$ is the corresponding maximized
likelihood from the fit. To take into account the systematic uncertainties discussed below, the
above likelihood in Eq.~(\ref{eq-3}) is convolved with a Gaussian function whose width
equals the total systematic uncertainty. Similarly, we estimate an upper limit for the branching
fraction of $\pcsb$ produced in $\yonetwos$ inclusive decays at 90\% C.L. by replacing
$N^{\rm fit}_{\rm sig}$ with $N^{\rm fit, UL}_{\pcsb}$ in Eq.~(\ref{brpcs}). The results are
listed in Table~\ref{ul}.

\begin{table}[htb]
\caption{The branching fractions and upper limits at 90\% C.L. for $\yonetwos$ inclusive decays
into $\pcs$ with $\pcs \to \lamjpsi$.}
\label{ul}
\begin{center}
\begin{tabular}{c   c}
\hline\hline
 Mode                                   & $\BR(\times 10^{-6})$ \\\hline
 $\yones \to \pcsa/\pcsabar +anything$  & $3.5 \pm 2.0 \pm 0.2$ \\
 $\ytwos \to \pcsa/\pcsabar +anything$  & $2.9 \pm 1.7 \pm 0.4$ \\
 $\yones \to \pcsb/\pcsbbar +anything$  & $<1.8$                \\
 $\ytwos \to \pcsb/\pcsbbar +anything$  & $<1.6$                \\
\hline\hline
\end{tabular}
\end{center}
\end{table}


Table~\ref{tab-sys} summarizes the systematic uncertainties in the determination of the
branching fractions and cross section. The uncertainties due to lepton identification are
2.0\% for the $\EE$ mode and 0.5\% for the $\MM$ mode from $\jpsi$ decay, contributing 1.4\% in
total. The uncertainty due to tracking efficiency is 0.35\% per track and added linearly.
The $\Lambda$ reconstruction uncertainties are estimated from a $\Lambda$ sample with a loose
selection. They are 4.0\% for $\yones$ decays, 3.6\% for $\ytwos$ decays, and 3.4\% for
continuum production, corresponding to the efficiency difference between data and MC. By
fitting to the $M_{\LL}$ and $M_{p\pi^{-}}$ distributions, we obtain the efficiencies of the
mass windows from data~($\eff^{\rm data}$) and signal MC simulation~($\eff^{\rm MC}$) in
$\yonetwos$ inclusive decays. We take the ratio of ${|\eff^{\rm data}_{\jpsi} -
\eff^{\rm MC}_{\jpsi}|}/{\eff^{\rm data}_{\jpsi}}$ as a conservative estimate of the systematic
uncertainty, and find 2.1\% (1.0\%) in $\yones$ [$\ytwos$] decays, and 2.0\% in continuum
production for the $\jpsi$ mass window. Similarly, we obtain 1.6\% (3.2\%) in $\yones$
[$\ytwos$] decays, and 2.7\% in continuum production for the $\Lambda$ mass window. The
uncertainties of modeling the final states in MC simulations are estimated by varying the mean
mass of the $\qqb$ system to $3.3~\gevcs$ ($2.5~\gevcs$) for $\yones$ [$\ytwos$] decays. We find
that the efficiency changes to be 1.8\% in $\yones$ decays, 1.7\% in $\ytwos$ decays, and 1.8\%
in continuum production, and take these values as the systematic uncertainties. The
uncertainties of the efficiencies of the ``no-$\pcs$ MC samples'' are estimated by studying the
variations in efficiencies across different MC samples with different accompanying particles,
such as $\yonetwos \to \jpsi \Lambda\bar{\Lambda} + \qqb$ and $\EE \to \jpsi
\Lambda\bar{\Lambda}  + \qqb$. We take the efficiency differences as the systematic
uncertainties, which are 2.3\% in $\yones$ decay, 3.5\% in $\ytwos$ decay, and 1.9\% in
continuum production. We use the Particle Data Group values ~\cite{PDG} for the uncertainties
in the branching fractions for $\jpsi\to \EE/\MM$ decays (1.1\%), $\Lambda \to p\pi^-$ decays
(0.8\%), and the $\ytwos$ to $\yones$ decays (6.1\%). The uncertainty in the total number of
$\yones$ [$\ytwos$] events in the data sample is 2.0\% (2.6\%)~\cite{y1sevnt, y2sevnt}. The
uncertainty in the integrated luminosity of each of the data samples is 1.4\% and they are highly
correlated, which cancels in the scale factor $f_{\rm scale}$. The statistical uncertainties of
the signal MC samples are 0.5\% in common. By varying the photon energy cutoff by $50~\mev$ in
the simulation of ISR, we determine the change of 1+$\delta_{\rm ISR}$ to be 0.01 and take
1.0\% to be a conservative systematic uncertainty in the cross section $\sigma(\EE \to \jpsillb
+ anything)$ at $\sqrt{s}$ = 10.52 $\gev$. We sum the uncertainties in quadrature, assuming
they are independent, and obtain total systematic uncertainties of 6.4\%, 9.5\%, and 6.2\% for
the measurements of $\yones$ decays, $\ytwos$ decays and continuum $\EE$ annihilation,
respectively.

\begin{table}[htbp]
\caption{Systematic uncertainties (\%) in the $\jpsillb$ production measurement.}
\label{tab-sys}
\centering
\begin{tabular}{c  c  c c }
\hline\hline
 Source                 & $\yones$ & $\ytwos$ & $\EE$ annihilation \\\hline
 PID                    & 1.4      & 1.4      & 1.4                \\
 Tracking               & 1.4      & 1.4      & 1.4                \\
 $\Lambda$ selection    & 4.0      & 3.6      & 3.4                \\
 $\jpsi$ mass window    & 2.1      & 1.0      & 2.0                \\
 $\Lambda$ mass window  & 1.6      & 3.2      & 2.7                \\
 Mean mass of $\qqb$ system & 1.8  & 1.7      & 1.8                \\
 Accompanying particle  & 2.3      & 3.5      & 1.9                \\
 Branching fractions     & 1.4      & 6.3      & 1.4               \\
 $N_{\yonetwos}$        & 2.0      & 2.6      & --                 \\
 Luminosity             & --       & --       & 1.4                \\
 MC sample statistics   & 0.5      & 0.5      & 0.5                \\
 $1+\delta_{\rm ISR}$   & --       & --       & 1.0                \\\hline
 Sum in quadrature      & 6.4      & 9.5      & 6.2                \\
\hline\hline
\end{tabular}
\end{table}

To determine the systematic uncertainties of the resonant parameters of the $\pcsa$ structure,
we change the following input parameters to the fit. The two-dimensional sidebands are shifted
by $\pm 1 \sigma$ in the resolutions. The exponential function in $f_{\rm SB}$ is replaced with
a second-order Chebyshev function. The bin width is changed from $1~\mevcs$ to $2~\mevcs$. The
mass resolution is varied by 10\%. The estimation of the systematic uncertainty in modeling the
no-$\pcs$ MC simulation is the same as described previously, and the decay mode
$\yonetwos \to \jpsi \Lambda \bar{\Lambda} + \qqb$ is also taken into account. The differences
between the nominal fit results and those from these fits are taken as the systematic
uncertainties, which are $0.6~\mevcs$ for the mass and $2.7~\mev$ for the width.


In conclusion, using Belle data samples, we report the first observation of $\lamjpsi$
production in $\yonetwos$ decays and $\EE$ continuum annihilation. We measure the branching
fractions to be $\BR[\yones\to \jpsillb + anything] = ( 36.9 \pm 5.3 \pm 2.4) \times 10^{-6}$
and $\BR[\ytwos \to \jpsillb + anything] = (22.3 \pm 5.7 \pm 3.1) \times 10^{-6}$, and the
cross section to be $\sigma(\EE \to \jpsillb + anything) = (90 \pm 14 \pm 6)~\fb$ at $\sqrt{s}
= 10.52~\gev$. We find a resonance-like peak in the $\lamjpsi$ invariant mass distribution in
the combined $\yonetwos$ data sample. The significance of the excess, assuming that it has the
same origin as the $\pcsa$ candidate~\cite{pcs} is $3.3\sigma$, including systematic
uncertainties. This is the first evidence of an exotic state in $\yonetwos$ decay. Furthermore,
the observation in $\yonetwos$ decay indicates a significantly different production mechanism than that of the LHCb evidence found in $\Xi_b^-$ decay. The branching fraction for $\pcsa$ production in inclusive $\yonetwos$ decay are
determined to be $\BR[\yones\to \pcsa/\pcsabar + anything] = (3.5 \pm 2.0 \pm 0.2) \times
10^{-6}$ and $\BR[\ytwos \to \pcsa/\pcsabar + anything] = (2.9 \pm 1.7 \pm 0.4) \times 10^{-6}$.
The mass and width of $\pcsa$ are measured to be $(4471.7 \pm 4.8 \pm 0.6)~\mevcs$ and $(22
\pm 13 \pm 3) ~\mev$, respectively. We determine upper limits on $\pcsb$ productions in the
$\yonetwos$ inclusive decays to be $\BR[\yones \to \pcsb/\pcsbbar +anything] \cdot \BR(\pcsb \to
\lamjpsi) <  1.8 \times 10^{-6}$ and $\BR[\ytwos \to \pcsb/\pcsbbar] +anything] \cdot \BR(\pcsb
\to \lamjpsi) <  1.6 \times 10^{-6}$.

\acknowledgments

This work, based on data collected using the Belle II detector, which was built and commissioned
prior to March 2019, was supported by
Higher Education and Science Committee of the Republic of Armenia Grant No.~23LCG-1C011;
Australian Research Council and Research Grants
No.~DP200101792, 
No.~DP210101900, 
No.~DP210102831, 
No.~DE220100462, 
No.~LE210100098, 
and
No.~LE230100085; 
Austrian Federal Ministry of Education, Science and Research,
Austrian Science Fund (FWF) Grants
DOI:~10.55776/P34529,
DOI:~10.55776/J4731,
DOI:~10.55776/J4625,
DOI:~10.55776/M3153,
and
DOI:~10.55776/PAT1836324,
and
Horizon 2020 ERC Starting Grant No.~947006 ``InterLeptons'';
Natural Sciences and Engineering Research Council of Canada, Compute Canada and CANARIE;
National Key R\&D Program of China under Contract No.~2022YFA1601903,
National Natural Science Foundation of China and Research Grants
No.~11575017,
No.~11761141009,
No.~11705209,
No.~11975076,
No.~12135005,
No.~12150004,
No.~12161141008,
No.~12475093,
and
No.~12175041,
and Shandong Provincial Natural Science Foundation Project~ZR2022JQ02;
the Czech Science Foundation Grant No.~22-18469S 
and
Charles University Grant Agency project No.~246122;
European Research Council, Seventh Framework PIEF-GA-2013-622527,
Horizon 2020 ERC-Advanced Grants No.~267104 and No.~884719,
Horizon 2020 ERC-Consolidator Grant No.~819127,
Horizon 2020 Marie Sklodowska-Curie Grant Agreement No.~700525 ``NIOBE''
and
No.~101026516,
and
Horizon 2020 Marie Sklodowska-Curie RISE project JENNIFER2 Grant Agreement No.~822070 (European grants);
L'Institut National de Physique Nucl\'{e}aire et de Physique des Particules (IN2P3) du CNRS
and
L'Agence Nationale de la Recherche (ANR) under Grant No.~ANR-21-CE31-0009 (France);
BMBF, DFG, HGF, MPG, and AvH Foundation (Germany);
Department of Atomic Energy under Project Identification No.~RTI 4002,
Department of Science and Technology,
and
UPES SEED funding programs
No.~UPES/R\&D-SEED-INFRA/17052023/01 and
No.~UPES/R\&D-SOE/20062022/06 (India);
Israel Science Foundation Grant No.~2476/17,
U.S.-Israel Binational Science Foundation Grant No.~2016113, and
Israel Ministry of Science Grant No.~3-16543;
Istituto Nazionale di Fisica Nucleare and the Research Grants BELLE2,
and
the ICSC – Centro Nazionale di Ricerca in High Performance Computing, Big Data and Quantum Computing, funded by European Union – NextGenerationEU;
Japan Society for the Promotion of Science, Grant-in-Aid for Scientific Research Grants
No.~16H03968,
No.~16H03993,
No.~16H06492,
No.~16K05323,
No.~17H01133,
No.~17H05405,
No.~18K03621,
No.~18H03710,
No.~18H05226,
No.~19H00682, 
No.~20H05850,
No.~20H05858,
No.~22H00144,
No.~22K14056,
No.~22K21347,
No.~23H05433,
No.~26220706,
and
No.~26400255,
and
the Ministry of Education, Culture, Sports, Science, and Technology (MEXT) of Japan;  
National Research Foundation (NRF) of Korea Grants
No.~2016R1-D1A1B-02012900,
No.~2018R1-A6A1A-06024970,
No.~2021R1-A6A1A-03043957,
No.~2021R1-F1A-1060423,
No.~2021R1-F1A-1064008,
No.~2022R1-A2C-1003993,
No.~2022R1-A2C-1092335,
No.~RS-2023-00208693,
No.~RS-2024-00354342
and
No.~RS-2022-00197659,
Radiation Science Research Institute,
Foreign Large-Size Research Facility Application Supporting project,
the Global Science Experimental Data Hub Center, the Korea Institute of
Science and Technology Information (K24L2M1C4)
and
KREONET/GLORIAD;
Universiti Malaya RU grant, Akademi Sains Malaysia, and Ministry of Education Malaysia;
Frontiers of Science Program Contracts
No.~FOINS-296,
No.~CB-221329,
No.~CB-236394,
No.~CB-254409,
and
No.~CB-180023, and SEP-CINVESTAV Research Grant No.~237 (Mexico);
the Polish Ministry of Science and Higher Education and the National Science Center;
the Ministry of Science and Higher Education of the Russian Federation
and
the HSE University Basic Research Program, Moscow;
University of Tabuk Research Grants
No.~S-0256-1438 and No.~S-0280-1439 (Saudi Arabia), and
Researchers Supporting Project number (RSPD2025R873), King Saud University, Riyadh,
Saudi Arabia;
Slovenian Research Agency and Research Grants
No.~J1-9124
and
No.~P1-0135;
Agencia Estatal de Investigacion, Spain
Grant No.~RYC2020-029875-I
and
Generalitat Valenciana, Spain
Grant No.~CIDEGENT/2018/020;
The Knut and Alice Wallenberg Foundation (Sweden), Contracts No.~2021.0174 and No.~2021.0299;
National Science and Technology Council,
and
Ministry of Education (Taiwan);
Thailand Center of Excellence in Physics;
TUBITAK ULAKBIM (Turkey);
National Research Foundation of Ukraine, Project No.~2020.02/0257,
and
Ministry of Education and Science of Ukraine;
the U.S. National Science Foundation and Research Grants
No.~PHY-1913789 
and
No.~PHY-2111604, 
and the U.S. Department of Energy and Research Awards
No.~DE-AC06-76RLO1830, 
No.~DE-SC0007983, 
No.~DE-SC0009824, 
No.~DE-SC0009973, 
No.~DE-SC0010007, 
No.~DE-SC0010073, 
No.~DE-SC0010118, 
No.~DE-SC0010504, 
No.~DE-SC0011784, 
No.~DE-SC0012704, 
No.~DE-SC0019230, 
No.~DE-SC0021274, 
No.~DE-SC0021616, 
No.~DE-SC0022350, 
No.~DE-SC0023470; 
and
the Vietnam Academy of Science and Technology (VAST) under Grants
No.~NVCC.05.12/22-23
and
No.~DL0000.02/24-25.

These acknowledgements are not to be interpreted as an endorsement of any statement made
by any of our institutes, funding agencies, governments, or their representatives.

We thank the SuperKEKB team for delivering high-luminosity collisions;
the KEK cryogenics group for the efficient operation of the detector solenoid magnet and IBBelle on site;
the KEK Computer Research Center for on-site computing support; the NII for SINET6 network support;
and the raw-data centers hosted by BNL, DESY, GridKa, IN2P3, INFN, 
and the University of Victoria.

\end{document}